\documentclass[prd,
showpacs,floatfix]{revtex4}
\usepackage{bm}
\usepackage{amssymb}
\usepackage{graphicx}
\usepackage{epstopdf}

\begin{document}
\title{Late--time Kerr tails: generic and non--generic initial data sets, ``up" modes, and superposition}
\author{Lior M. Burko$^{1,2}$ and Gaurav Khanna$^{3}$}
\affiliation{
$^1$Department of Physics, University of Alabama in Huntsville, Huntsville, Alabama 35899, USA \\
$^2$Center for Space Plasma and Aeronomic Research, University of Alabama in Huntsville, Huntsville, Alabama 35899, USA
\\
$^3$Physics Department, University of Massachusetts at Dartmouth, N. Dartmouth, Massachusetts 02747, USA }
\date{October 18, 2010}
\begin{abstract}
Three interrelated questions concerning Kerr spacetime late--time scalar--field tails are considered numerically, specifically the evolutions of generic and non--generic initial data sets, the excitation of ``up" modes, and the resolution of an apparent paradox related to the superposition principle. We propose to generalize the Barack--Ori formula for the decay rate of any tail multipole given a generic initial data set, to the contribution of any initial multipole mode. Our proposal leads to a much simpler expression for the late--time power law index. Specifically, we propose that the late--time decay rate of the $Y_{\ell m}$ spherical harmonic multipole moment because of an initial $Y_{\ell' m}$ multipole is independent of the azimuthal number $m$, and is given by $t^{-n}$, where $n=\ell'+\ell+1$ for $\ell<\ell'$ and $n=\ell'+\ell+3$ for $\ell\ge\ell'$.
We also show explicitly that the angular symmetry group of a multipole does not determine its late--time decay rate. 
\end{abstract}
\maketitle

\section{Introduction and Summary}\label{sec:int}

Perturbations of black holes decay first with complex frequencies, known as the quasinormal modes of of the black hole, taken over at late times by the tails of the perturbation field--- decaying as power--laws of time. The decay rate of the late--time tails in the spacetime of spinning black holes has been the subject of much debate and some confusion. 
Much of the debate in the literature was focused on the late--time decay rate of an initially pure azimuthal hexadecapole ($\ell'=4$, $m=0$) scalar field perturbation. All authors agree that at late times the field is dominated by its monopole moment, and that the decay rate would be according to an inverse power of time, but different claims were made as to the value of the power--law exponent. Specifically, there were claims for decay rate along fixed Boyer--Lindquist $r$ according to $t^{-3}$ \cite{burko-khanna03}, $t^{-5}$ \cite{hod,poisson}, and even $t^{-5.5}$ \cite{krivan}. 

In the last couple of years the behavior of the power--law tails has been clarified in a number of  papers \cite{GPP,TKT,burko-khanna09} (and the reader is referred to the detail therein), yet there are some remaining interesting detail waiting to be unveiled. In particular, it was shown in \cite{burko-khanna09} that the disagreement between the $t^{-3}$ and $t^{-5}$ behaviors can be attributed to the use of different initial data sets (see also \cite{TKT}), but not to the use of different slicing conditions as was suggested in \cite{GPP}. Most importantly, Boyer--Lindquist slicing is not unique in terms of the resulting tails as had been argued, but rather belongs to an equivalency class of slicing conditions (including also ingoing Kerr slicing), all having similar tail structure properties. 

For the evolution of scalar fields on a fixed background of a Kerr black hole, the value of the power-law index, namely the exponent $n$ in the late-time decay rate $t^{-n}$, was given by Hod \cite{hod} as 
$n= \ell'+m+1$ for even $\ell'-m>2$, n=$\ell'+m+2$ for odd $\ell'-m>2$, and $n=2\ell'+3$ for $\ell'-m<2$, where $\ell'$ is the multipole moment of the initial perturbation and $m$ is its azimuthal number. This expression for $n$ is rather complicated and notably depends on the azimuthal number $m$. We argue that the added tacit requirement that it is the decay rate of the slowest decaying mode that is of usual interest masks the inherent simplicity of $n$ in the Hod formula. Specifically, we propose that the decay rate of the $\ell$ mode because of an initial $\ell'$ mode is independent of $m$, and is given by $n=\ell'+\ell+1$ for $\ell<\ell'$ and $n=\ell'+\ell+3$ for $\ell\ge\ell'$. Notice that this proposal is consistent with Hod's formula and with \cite{burko-khanna09} when they are applicable, and generalizes them also to all other cases. It is therefore a proposal for the general formula for the decay rate of any multipole moment of the tail because of any multipole moment of the initial data. It is striking that this more general formula is simpler than the formula for the decay rate of the least damped mode, that was the question that \cite{hod} was addressing. We believe that this simple, general formula will provide insight into a more complete understanding of the tail phenomenon in spinning black hole spacetimes.


Our general formula allows up to address some questions left open also in the analysis of Barack and Ori \cite{barack-ori99}. Barack and Ori studied analytically the late--time tails for generic families of initial data (only assuming compactly supported outgoing initial pulses). Instead of carrying the analysis in the frequency domain ---which turns out to be complicated on a Kerr background because the frequency dependence of the spheroidal harmonics implies that separation of the two angular variables depends on the frequency--- Barack and Ori analyzed the evolution of perturbations in the time domain. Barack and Ori found the decay rate of the $\ell$ multipole given generic initial data, which they considered to be data in which all multipoles are present. In fact, because the smallest multipole contribution to the initial data turns out as we show below to dominate the decay rate of the $\ell$ multipole at late times, Barack and Ori found the decay rate of an even (odd) mode $\ell$ excites by the scalar field monopole (dipole) mode of the initial data. The Barack--Ori formula for the decay rate of the $\ell$ multipole has not been confirmed by numerical simulations, which we do below for the first time. Our proposed formula also generalizes it to any initial multipole, not just the monopole (or dipole) as in \cite{barack-ori99}.

Consider next the question of the decay rate of a certain multipole moment $\ell$. In the Schwarzschild spacetime the decay rate depends only the the value of $\ell$, specifically it is $t^{-n}$ where $n=2\ell+3$. Most notably, the decay rate is independent of the history of the mode. This is no longer the case in the Kerr spacetime: the decay rate of a multipole $\ell$ does depend on how that mode came into existence, specifically, it depends on both $\ell,\ell'$. In the Schwarzschild case the dependence is only on $\ell$ because of the degeneracy $\ell'=\ell$ and the absence of mode couplings (``spherical harmonics are 
eigenfunctions of the wave operator") . Therefore, in the Schwarzschild case one may say that ``a multipole is a multipole," and consequently its decay rate is intimately linked to its angular distribution and the associated symmetry group.   Specifically, in Schwarzschild there is a one--to--one relation of the symmetry group of the mode in question and the mode's late--time decay rate, such that the symmetry determines the time evolution of the mode. Previous studies of tail behavior in Kerr have suggested that it is not correct to say in Kerr that ``a multipole is a multipole,"  although the authors are not aware of explicit statements in this regard. We bring here new evidence that that is indeed the case, and the tail decay rate has nothing to do with the symmetry group of the mode in question.  This insight may be helpful in the understanding of cases where more than one channel contributes to a mode of a certain $\ell$ value, although each channel contributes so that the decay rate is different, in what might be construed as an apparent violation of the superposition principle.  The realization that in Kerr one may no longer correctly argue that ``a multipole is a multipole" completely resolves this apparent violation.

The organization of this Paper is as follows: In Section \ref{sec:code} we describe the numerical code and the type of initial data sets that we use.  Specifically, we describe the code development that was necessary in order to support our proposal for the tail decay-rate formula. In this Paper we discuss three related questions: First, in Section \ref{sec:generic} we address the difference in the time evolution of late--time tails between generic and non--generic initial data sets. Generic initial data sets were first considered by Barack and Ori \cite{barack-ori99}, but their results have not been confronted with numerical simulations. In addition to verifying the results of \cite{barack-ori99}, we also discuss the meaning of genericity of initial data sets. As we demonstrate, not any mixture of all multipole modes is in fact generic, and one needs to exclude mixtures that are the outcome of the evolution of non-generic initial data, although such mixtures exhibit similar properties to mixtures that represent {\em bona fide} generic data sets. Then, in Section \ref{sec:up}, we consider an aspect of late--time tails that has not been studied before, specifically the decay rate of excited higher multipole modes, which we call ``up"--excited modes, namely the case $\ell>\ell'$. The decay rate of ``up" excited modes generalized the Barack--Ori result in \cite{barack-ori99} to some interesting cases: the case of non--generic initial data in which (only) the lowest dynamically allowed mode is not present, and the case in which only one multipole mode is present in the initial data. 

\section{Numerical code}\label{sec:code}

\subsection{Initial data sets}

Linearized perturbations of Kerr black holes are described by Teukolsky's master equation, given in Boyer--Lindquist coordinates for a scalar field ($s=0$) $\psi$, known as the Teukolsky function, by 
\begin{eqnarray}\label{teuk}
&-&\left[\frac{(r^2+a^2)^2}{\Delta}-a^2\,\sin^2\theta\right]\,\partial_{tt}\psi-\frac{4Mar}{\Delta}\,\partial_{t\varphi}\psi\nonumber\\
&+&\,\partial_r\left(\Delta\,\partial_r\psi\right)+\frac{1}{\sin\theta}\,\partial_{\theta}\left(\,\sin\theta\,\partial_{\theta}\psi\right)\nonumber\\
&+&\left(\frac{1}{\,\sin^2\theta}-\frac{a^2}{\Delta}\right)\,\partial_{\varphi\varphi}\psi=0\, ,
\end{eqnarray}
where $M,a$ are the black hole's mass and spin angular momentum per unit mass, respectively, and  the horizon function $\Delta=r^2-2Mr+a^2$.

To solve Eq.~(\ref{teuk}) numerically, we define, following \cite{KLP}, 
$b(r,\theta):=(r^2+a^2)/\Sigma$ where $\Sigma^2=(r^2+a^2)^2-a^2\,\Delta\,\sin^2\theta$. We next define the `momentum' $\Pi$ of the field $\phi$ defined by $\psi=e^{im\varphi}\,\phi$, according to
$$\Pi:=\frac{\,\partial\phi}{\,\partial t}+b(r,\theta)\,\frac{\,\partial\phi}{\,\partial r_*}\, $$
where $r_*$ is the regular Kerr spacetime `tortoise' coordinate defined by $\,dr_*/\,dr=(r^2+a^2)/\,\Delta$.

Our code is a 2+1D first--order code for the time evolution of modes $m$ of $\phi,\Pi$ given two independent functions of two variables at $t=0$ as initial data, based on \cite{KLP}. The convergence order is taken to be second--order in the radial and temporal directions, and sixth order in the angular direction (although some of the higher--$\ell$ results were obtained with tenth--order angular operators). The standard grid resolution we use is $64,000$ grid points in the radial direction, and $64$ angular grid points (which we denote as $64K\times 64$), unless stated otherwise, taking the temporal step to be the largest step consistent with the Courant condition. 

In practice, we choose as initial data $\phi(r,\theta)|_{t=0}=0$ and $\Pi(r,\theta)|_{t=0}=f(r)\,P_{\ell}(\,\cos\theta)$, where the radial function $f(r)$ is chosen to be a gaussian $f(r)=(1/\sqrt{2\pi\sigma^2})\,\exp[{-(r_*-r_{*\; 0})^2/(2\sigma^2)}]$ with $r_{*\;0}=25M$ and $\sigma=6M$. The outer and inner boundaries are placed at $r_*^{\rm boundary}=\pm 800M$, which allows us to integrate to $t=1,500M$ at $r_{*\;0}$ without seeing boundary reflection effects. We choose the Kerr parameter $a=0.995M$. We use quadrupole or octal precision floating--point arithmetic throughout according to need. 

We emphasize that our choice of initial data ---based on the field and its conjugate momentum--- is in accord with other common choices. Specifically, when we choose a single value of $\ell$, our choice of initial data is a pure multipole moment, equivalent to choices made based on the field and its time derivative. There is no ambiguity in the definition of a pure multipole \cite{burko-khanna09}.

\subsection{Code Improvements}

Since the goal of this work is to investigate the late-time decay rate of higher multipole modes of Kerr black holes a number of improvements to our time-domain Teukolsky equation code were necessary. This is because generating accurate numerical data in the context of higher multipole modes involves a number of challenges. Firstly, these simulations need to be rather long --- this is because typically the observed field exhibits an exponentially decaying oscillatory behavior (``quasi- normal ringing'') in the initial part of the evolution and only much later this transitions over to a clean power-law decay. One needs to wait for these oscillations to dissipate away. For higher multipoles this phase lasts much longer, thus requiring significantly longer evolutions. Secondly, because each multipole has a different decay rate (which increases with an increase in $\ell$) at late times one ends up with numerical data in which different multipoles have widely different amplitudes (often 30 -- 40 orders of magnitude apart!). For this reason, not only does the numerical solution scheme have to be high-order (to reduce the discretization errors to the required levels) but it also requires high-precision floating-point numerical computation (due to the large range of amplitudes involved). 

Below we list the major improvements made since the publication of \cite{burko-khanna09} to our Teukolsky equation evolution code:

\begin{enumerate}

\item {\em Higher-Order Algorithm:} As mentioned above, we require a higher-order numerical evolution scheme to keep the discretization errors to sufficiently low levels, otherwise these can overwhelm the numerical data associated to the weakly excited higher multipole modes of interest. It turns out that it is sufficient that only the angular differentiation (i.e. $\theta$-derivatives of the field) be implemented using a higher-order numerical stencil. The temporal and the radial direction related operations can simply stay 2nd-order and such a mixed approach yields sufficiently good results \cite{burko-khanna09}. For this reason, in this work we choose the finite-difference angular differentiation operator to be 10th-order accurate and leave the rest of the numerical scheme as a standard 2nd-order Lax-Wendroff algorithm. 

\item {\em Higher-Order Projection Integrals:} We also require a matching 10th-order accuracy in mode projection integrals to prevent numerical integration errors from overwhelming the higher multipole numerical data. 

\item {\em Higher-Order Numerical Precision:} As pointed out above, we also require high-numerical precision in our computations. In particular, double, quadruple and octal precision may be required depending upon the details of Kerr tails simulation being attempted. This is simply to prevent roundoff error from overwhelming the weakly excited modes of interest -- especially important for ``up" excited higher multiple modes! We made use of quadruple precision (in some cases, octal precision) floating-point accuracy in this work. Now, very few current computer processors support full quadruple-precision 
(128--bit) datatype and operations. And to the best of our knowledge, no compute hardware natively supports octal-precision (256--bit) arithmetic. Thus, we implemented such high-precision computations in software using publicly available software libraries. This made our work particularly challenging computationally, because with each such increase in numerical precision, our code's overall performance drops by an order-of-magnitude.

\item {\em Code Parallelism and HPC Hardware:} In order to perform the numerical simulations in a reasonable time-frame, we made use of message-passing (MPI) based parallelism in our code and also used sophisticated HPC hardware. In particular, domain-decomposition based parallelism was implemented in the code and IBM Cell BE and Power7 systems with specialized optimizations \cite{GK_2010} were used to perform these long evolutions in a reasonable time-frame. 

\item {\em Other Enhancements:} We also enhanced the code further to be able to evolve non-axisymmetric mode computations (both odd and even $m$-modes). The even $m$-mode case was straightforward to implement, since these simply require some additional terms in the evolution equation. For the odd $m$-mode case we used the following procedure to simplify our implementation: Instead of evolving the scalar field $\psi$ itself, we transformed the Teukolsky equation to evolve the field $\psi / \,\sin \theta$. This simple transformation enabled us to reuse several of our even $m$-mode code algorithms (such as the higher-order angular boundary condition imposition) and also avoid the introduction of non-polynomial functions of the code's angular grid variable, i.e., $\cos \theta$.

\end{enumerate}

\section{Numerical evolution of generic initial data}\label{sec:generic}

We first show for the first time numerical tests of the Barack--Ori formula \cite{barack-ori99}. Then, in Section \ref{sec:up} we show how the Barack--Ori formula is a particular case for ``up" excitation of modes. Barack and Ori noted that in the generic case of black hole perturbations, all dynamically allowed modes are present. Specifically, in the case of scalar field perturbations, all modes, starting with the monopole ($\ell=0$) evolve.  The decay rate of a multipole $\ell$ with azimuthal number $m$ is given by 
$t^{-n}$ where 
\begin{equation}\label{b-o}
n_{\ell}=\ell+|m|+3+q\, ,
\end{equation}
where $q=0,1$ if $\ell+m$ is even or odd, respectively. Notably, the Barack--Ori formula depends explicitly on the azimuthal number $m$, and like the Hod formula includes a certain complexity that has to do with whether $\ell+m$ is even or odd. 

The meaning of ``generic" initial data is that the relative amplitudes of the various multipoles and their radial profiles are not fine tuned. Consider, e.g., the following initial data set, $\phi(r,\theta)|_{t=0}=0$ and $\Pi(r,\theta)_{t=0}=A\,\sin^2\theta\,f(r)$, where $f(r)$ is any localized radial function, say a gaussian. These initial data include non-vanishing monopole and quadrapole moments. We evolved these data numerically, and found that as expected the late--time tail includes all even multipoles, and is dominated by the monopole field that decays according to $t^{-3}$ at late times. The quadrapole projection of the late--time field decays according to $t^{-5}$, in accordance with the Barack--Ori formula. Other choices of initial data sets and multipoles are also found to be in agreement with the Barack--Ori formula. As far as the present authors are aware, this is the first direct numerical confirmation of the Barack--Ori formula. 

Let us now consider the following scenario: one starts with initial data as above, and evolves the field to a certain value of time, say to $t=T$. The field at $t=T$ is a complicated mixture of modes, each with its own radial profile. Now consider a different initial value problem, in which at $t=0$ only a pure multipole, say the quadrapole moment, is non-vanishing, and evolve these initial data to $t=T$. These pure mode initial data also evolve to a complicated mixture of multipole moments at $t=T$, and at first look one does not notice much qualitative difference between the fields at $t=T$. However, treating the fields at $t=T$ as new initial value  problems, further time evolution leads to different decay rates at late times. Specifically, the generic data lead to quadrapole projection that drops off like $t^{-5}$, and the pure data lead to  quadrapole projection that drops off like $t^{-7}$. Considering only the values at $t=T$ as the initial data sets, what is the fundamental difference between the two sets, that leads to very different late--time evolutions? Even though either set appears at $t=T$ to be a set of amplitudes with weighted ratios for the various multipole moments (as functions of the radial coordinate), the set describing the pure mode evolution is made at $t=T$ of a very carefully chosen set of ($r$ dependent) amplitudes. In fact, it is the outcome of fine--tuned evolution of a pure mode of the original initial value problem. Examination of the data set at $t=T$ does not reveal anything qualitative different about it: it is the specific amplitude ratios and radial profiles that make it non-generic. 

\begin{figure}
 \includegraphics[width=3.4in]{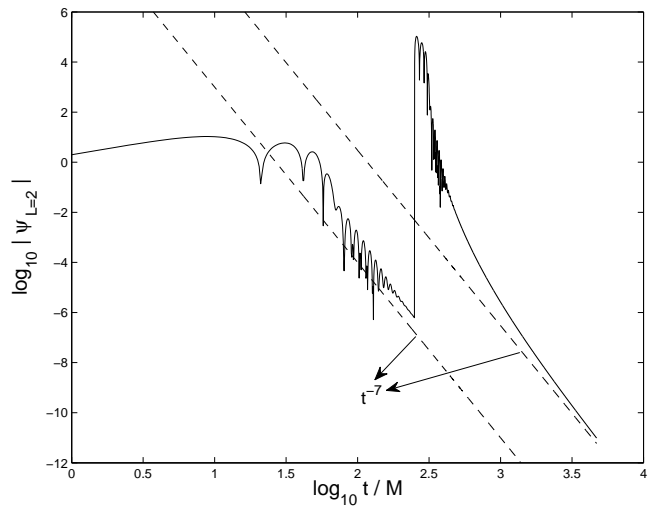} 
\caption{The quadrapole projection of the full field as a function of time, for initial data of a pure quadrapole field, then injected at $T=250M$ with a pure quadrapole field. The late--time field behavior is computationally found to be given by $\psi\sim t^{-7.05}$. The power law index, and all the indices in this Paper, were calculated by extrapolating the local power-law index to infinite time.}
\label{wave0}
\end{figure}

\begin{figure}
 \includegraphics[width=3.4in]{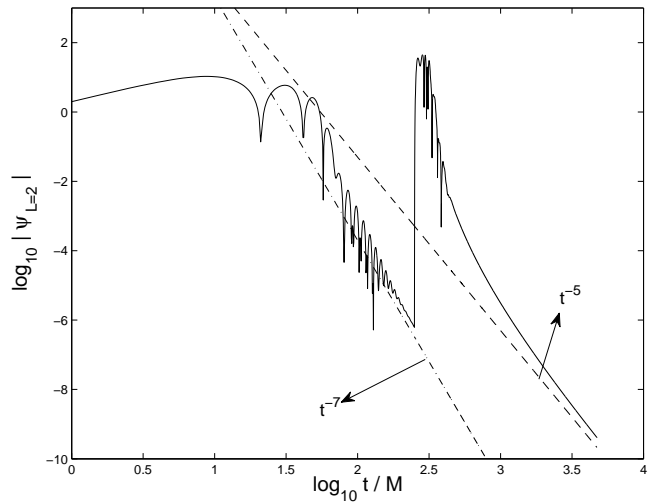} 
\caption{The quadrapole projection of the full field as a function of time, for initial data of a pure quadrapole field, then injected at $T=250M$ with a pure monopole field. The late--time field behavior is computationally found to be given by $\psi\sim t^{-5.03}$.}
\label{wave1}
\end{figure}

One may test these ideas by injecting at $t=T$ another field, say that of a pure mode. Start at $t=0$ with a pure quadrapolar field.   At $t=T$ the projection of the field is a mixture of all even multipoles. Then, at $t=T$, inject either a pure monopole or a pure quadrapole with some radial profile. The multipolar content of the field after the injection is changed in that the relative multipolar amplitudes changed (as functions of the radius). When the injected field is quadrapolar, the late--time decay rate of the quadrapole moment is $t^{-7}$, as should be expected from the fact that we have a superposition of two linear pure mode evolutions (which a certain time delay between them) (Fig.~\ref{wave0}). When the injected field is a monopole, the late--time decay rate of the quadrapole moment is $t^{-5}$ (Fig.~\ref{wave1}).  The addition of a monopole field (without fine tuning its amplitude and radial profile) makes the multipolar content of the total field generic. One may pose the following question: When the injected field is a quadrapole, its time evolution will excite a monopole. Why in the presence of this excited monopole the quadrapole projection of the field decays like $t^{-7}$, whereas in the presence of an injected monopole the decay rate is the slower $t^{-5}$? There is a fundamental difference between the two cases: the excited monopole is not an arbitrary field: it is carefully chosen by the dynamics of the problem, and may be viewed as a fine tuned field; the injected monopole is arbitrary, and is not fine tuned to lead to a different late--time decay rate. 

\section{Excitation of ``up" modes}\label{sec:up}

Most interest has been devoted to finding the decay rate of the slowest decaying mode, for obvious reasons. Specifically, the slowest decaying mode determines the late--time behavior of the full field. As lower multipole modes typically have a slower decay rate, most interest has naturally focused on the excitation of ``down" modes, specifically the excitation of lower multipole modes starting with higher multipole modes. 

Here, we are interested in the converse, i.e., in the excitation and decay rate of ``up" modes, i.e.,  the excitation and decay rate of a higher multipole mode than the one initially excited. Such modes are typically sub-dominant, and do not dominate the late--time behavior of the full field. Our interest in the ``up" modes is therefore only tangentially related to the question of the decay rate of the full field. Instead, it is the gaining of more insight towards a more complete understanding of the dynamics and mode coupling properties thereof we are interested in here. 
Denoting the initially excited {\em pure} multipole by $\ell'$, and the multipole moment of the field whose excitation and decay are is of interest by $\ell$ (so that for ``up"-modes $\ell>\ell'$ and for ``down"-modes $\ell<\ell'$), 
we propose that the late--time decay rate for excited $\ell,m$ modes along a $r={\rm const}$ worldline, is given by $t^{-n}$ where 
\begin{equation}\label{prop}
_{\ell'}n_{\ell}=\left\{
\begin{array}{cc}
\ell'+\ell+1  &  {\rm for}\;\;  \ell<\ell'  \\
\ell'+\ell+3  & {\rm for}\;\;  \ell\ge\ell'       
\end{array}
\right. \, .
\end{equation} 
Notably, this formula is independent of $m$. Specifically, for all (allowed) values of $m$ the decay rate of the $\ell$ excited multipole because of an initial $\ell'$ is the same. This formula is simpler than either the Hod or the Barack--Ori formulas, but is fully consistent with them in their domain of applicability. Specifically, when $\ell'=0,1$, Eq.~(\ref{prop}) coincides with the Barack--Ori formula (\ref{b-o}). In the Schwarzschild case there is no mode coupling, $\ell=\ell'$, and therefore $n=\ell'+\ell+3=2\ell+3$. The numerical evidence described below supports Eq.~(\ref{prop}) in all cases we have checked. 

\subsection*{Azimuthal modes}

To test our proposal ---and in particular the $m$-independence of the decay rate, which may be construed as a statement on the generally $m$-dependent Green's function--- we focus now on pure multipole initial data sets. 

\begin{figure}
 \includegraphics[width=3.4in]{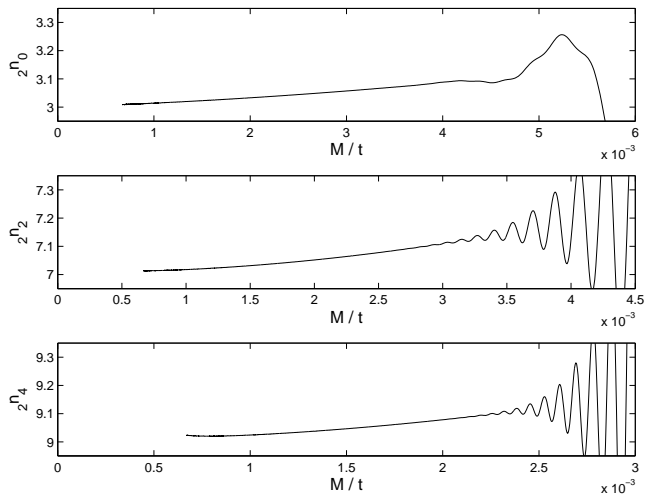} 
\caption{The $\ell=0,2,4$ multipole projections, for initial data of a pure quadrupole field as functions of time. The late--time field behavior is computationally found to be given by $_2\psi_0\sim t^{-3.005}$, $_2\psi_2\sim t^{-7.009}$, and $_2\psi_4\sim t^{-9.01}$, respectively.}
\label{fig41}
\end{figure}

Letting the initial $\ell'=2$, we project the multipoles $\ell=0,2,4$, and the local power indices are shown in Fig.~\ref{fig41}. Of particular interest here is the excitation of the azimuthal, hexadecapole mode, $\ell=4$. In this case, $\ell'=2,\ell=4$, so that our proposal is that $n=2+4+3=9$. In our numerical simulations, we have found the value of $_2n_4=9.01$, with uncertainly in the last figure. 

\begin{figure}
 \includegraphics[width=3.4in]{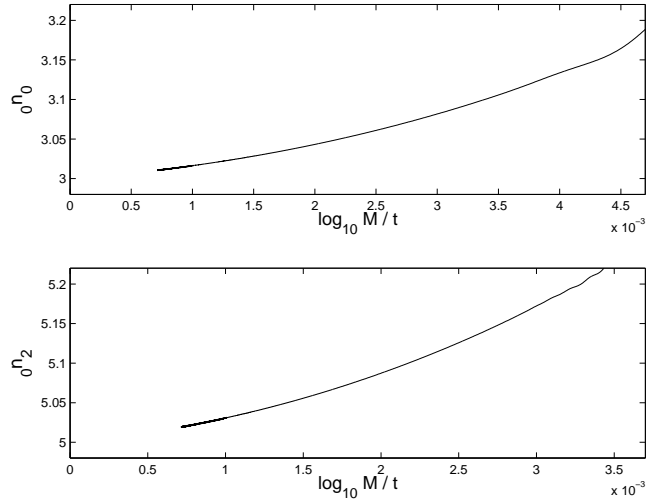} 
\caption{The local power indices $_0n_0$ (upper panel) and $_0n_2$ (lower panel) as functions of the time $t$.}
\label{lpi_002}
\end{figure}

\begin{figure}
 \includegraphics[width=3.4in]{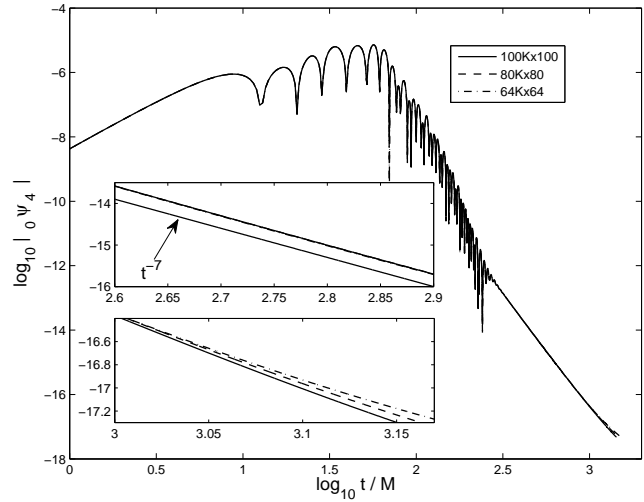} 
\caption{The $\ell=4$ multipole projections, for initial data of a pure monopole field as functions of time. Shown are the fields with three different grid densities, $64K\times 64$ (dash--dotted curve), $80K\times 80$ (dashed curve), and $100K\times 100$ (solid curve). The inserts show the same for segments of the full data, in addition to a reference curve $\sim t^{-7}$.}
\label{psi_04}
\end{figure}

\begin{figure}
 \includegraphics[width=3.4in]{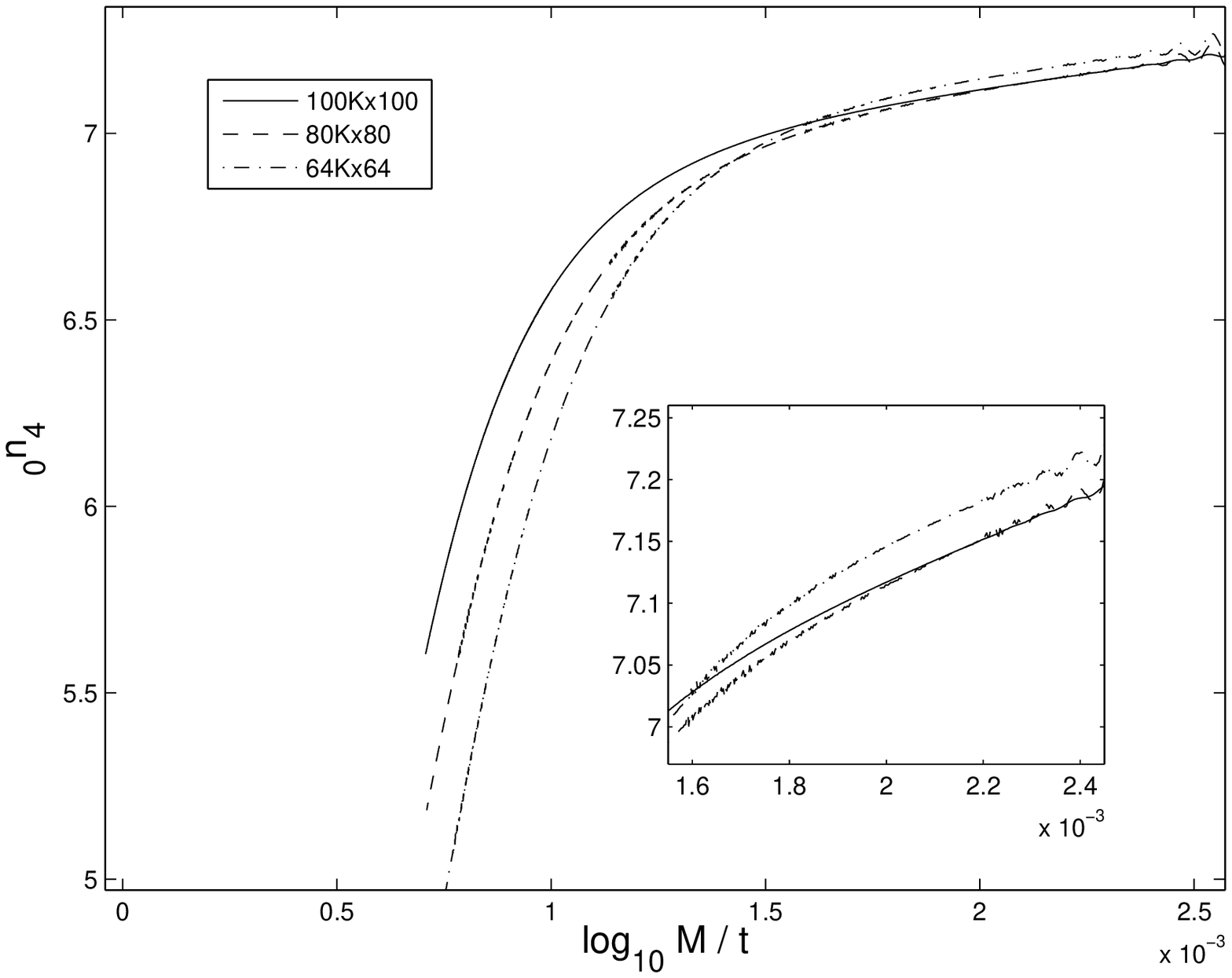} 
\caption{The local power index $_0n_4$ as a function of the time $t$ for three grid resolutions, $64K\times 64$ (dash--dotted curve), $80K\times 80$ (dashed curve), and $100K\times 100$ (solid curve).}
\label{lpi_04}
\end{figure}

Starting with initial data for a pure $\ell'=0$ monopole mode, we project the $\ell=0,2$ and $4$ modes. The local power indices $_0n_0$ and $_0n_2$ are shown in Fig.~\ref{lpi_002}.  
The hexadecapole mode $\ell=4$, $_0\psi_4$, is shown in Fig.~\ref{psi_04}. Figure \ref{psi_04} suggests that our grid density is not sufficiently high to obtain accurately the very late time hexadecapole projection to the required accuracy. Indeed, as Fig.~\ref{psi_04} shows, the tail of the field ``curves up" at very late times. However, increasing the grid density appears to straighten up the field, from which we infer that this curving is a numerical artifact. We attribute the problem to our calculation of a mode that is a second--order excited ``up" mode, i.e., the $\ell=4$ excited mode from an initial $\ell'=0$ mode. 
The practical problems in determining the decay rate accurately are accentuated by Fig.~\ref{lpi_04}, which displays the local power indices for three grid densities. Figure \ref{lpi_04} suggests that indeed a higher resolution simulation would be successful at determining the late--time decay rate with sufficient accuracy. Our results do suggest that $_0n_4=7$ (notice, this is a numerical result with one significant figure, not an exact value!), as indicated by the reference curve in Fig.~\ref{psi_04}, but any determination of the decay rate at higher accuracy than that would require higher resolution simulations.

\begin{table}[htdp]
\caption{Scattering matrix for azimuthal modes. Late--time power law indices $_{\ell'}n_{\ell}$ for the $\ell$--projections of the fields for pure azimuthal ($m=0$) mode initial data sets, $\ell'=0,1,2,3,4$. This table includes both ``down" and ``up" excitations. Asterisks relate to kinematically disallowed states. All figures are significant.}
\begin{center}
\begin{tabular}{||c||c|c|c|c|c|c||}
\hline
Initial  & Projected & Projected & Projected & Projected & Projected &Projected\\
$\ell'$ mode & $\ell=0$ & $\ell=1$ & $\ell=2$ & $\ell=3$ & $\ell=4$  & $\ell=5$ \\
\hline
0 & 3.007 & * & 5.01 & * & 7 & *  \\
1 &  * & 5.002 & *  &  7.003  &  *  & 9  \\
2 & 3.005 & * & 7.009 & * & 9.01 & *  \\
3& * & 5.002  & * & 9.009 & * &  11.008 \\
4 &  5.01 & * & 7.002 & * & 11.008 & * \\
\hline
\end{tabular}
\end{center}
\label{default}
\end{table}%

\begin{figure}
 \includegraphics[width=3.4in]{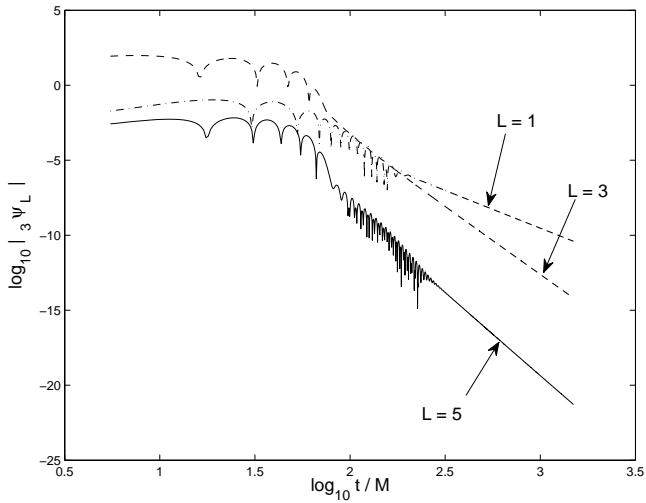} 
\caption{The $\ell=1,3,5$ multipole projections, for initial data of a pure octupole field as functions of time. The late--time field behavior is computationally found to be given by $\psi_1\sim t^{-5.002}$, $\psi_3\sim t^{-9.009}$, and $\psi_5\sim t^{-11.008}$, respectively.}
\label{fig32}
\end{figure}

\begin{figure}
 \includegraphics[width=3.4in]{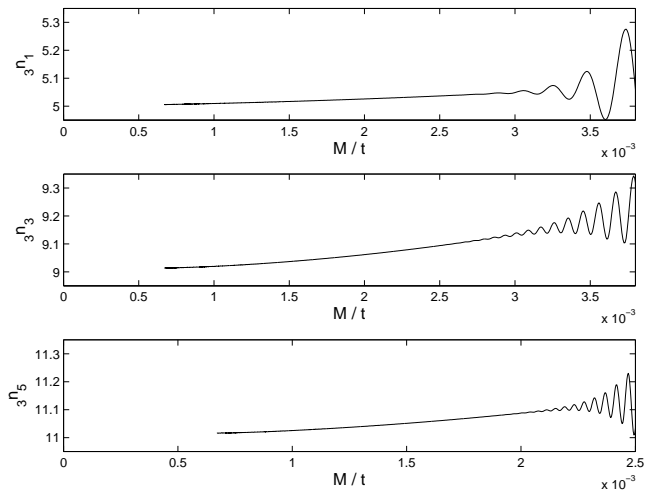} 
\caption{The local power indices of the $\ell=1,3,5$ multipole projections as functions of time, for initial data of a pure octupole field (same data as in Fig.~\ref{fig32}).}
\label{fig31}
\end{figure}

We have also tested our proposal with odd modes. Specifically, Let $\ell'=3$ and study the late time tail for the modes $\ell=1,3,5$. Figure \ref{fig32} shows the projections $\ell=1,3,5$ as functions of time, and figure \ref{fig31} displays the local power indices for these modes as functions of time. 

\begin{figure}
 \includegraphics[width=3.4in]{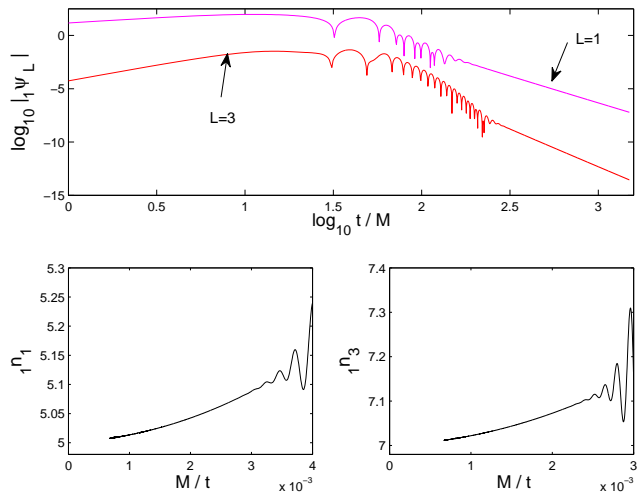} 
\caption{The dipole ($\ell=1$) and octupole ($\ell=3$) projections, $_1\psi_3$ and $_1\psi_3$, correspondingly (upper panel),  and their local power indices, $_1n_1$ and $_1n_3$, respectively (left and right lower panels), as functions of time, for initial data of a pure dipole ($\ell'=1$) field.}
\label{psi11n3}
\end{figure}

\begin{figure}
 \includegraphics[width=3.4in]{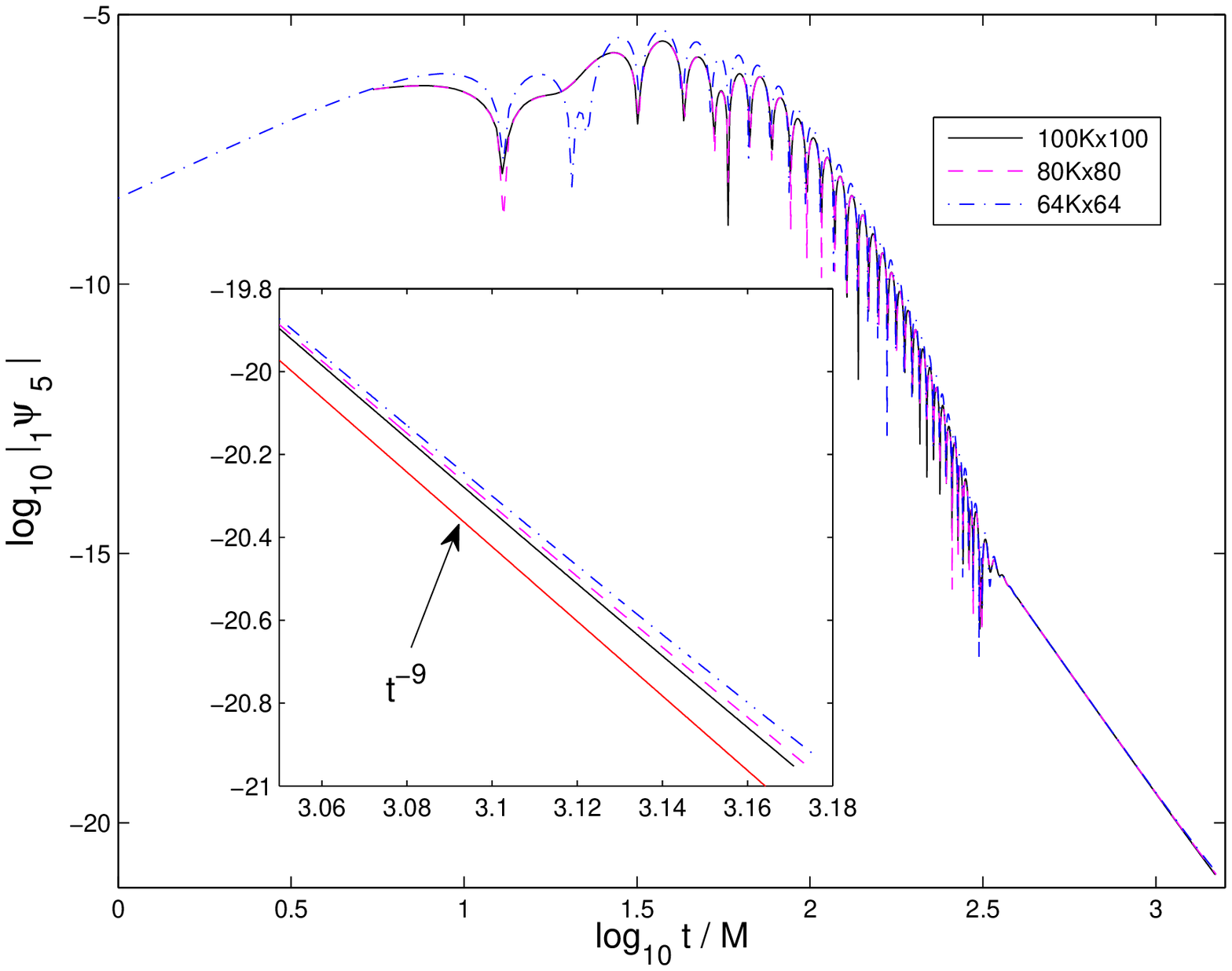} 
\caption{The $\ell=5$ multipole projections, for initial data of a pure monopole field as functions of time. Shown are the fields with three different grid densities, $64K\times 64$ (dash--dotted curve), $80K\times 80$ (dashed curve), and $100K\times 100$ (solid curve). The inserts show the same for segments of the full data, in addition to a reference curve $\sim t^{-9}$.}
\label{psi_15}
\end{figure}

\begin{figure}
 \includegraphics[width=3.4in]{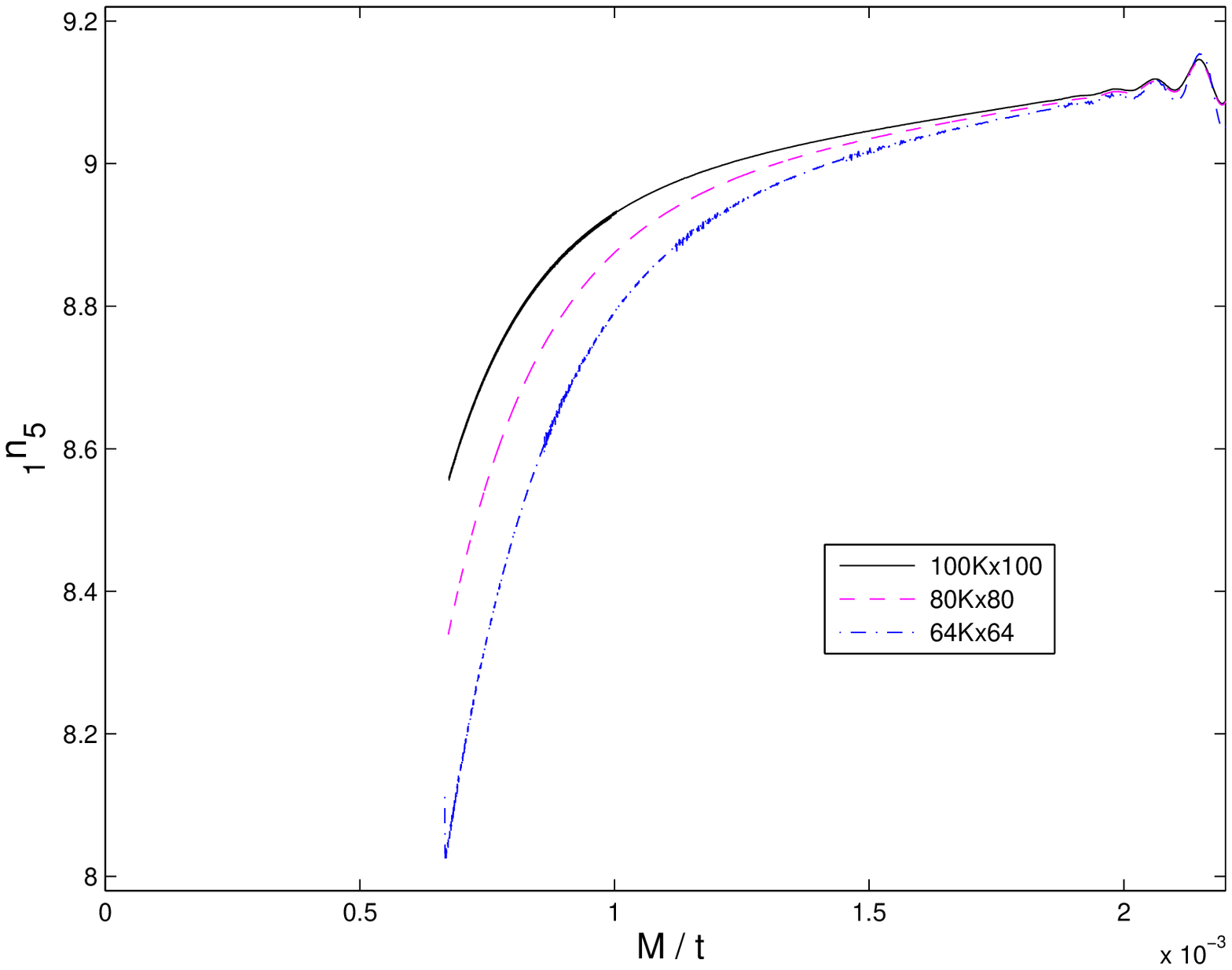} 
\caption{The local power index $_1n_5$ as a function of the time $t$ for three grid resolutions, $64K\times 64$ (dash--dotted curve), $80K\times 80$ (dashed curve), and $100K\times 100$ (solid curve).}
\label{lpi_15}
\end{figure}

Starting with $\ell'=1$, we project the dipole and octupole modes and their respective local power indices in Fig.~\ref{psi11n3}. 
We encounter the same problem finding the decay rate for $\ell=5$ --- i.e., find $_1\psi_5$ and $_1n_5$ --- as we did when we we calculated $_0\psi_4$ and $_0n_4$ (Figs.~\ref{psi_15} and \ref{lpi_15}). Again, the second--order ``up"--excited mode requires high resolution simulations. 

Our proposal (\ref{prop}) naturally extends the Barack--Ori formula (\ref{b-o}) to cases where the initial data do not include the monopole ($\ell'=0$) mode. Recall that the Barack--Ori formula considers the case of generic initial data, which means a mixture of all multipole modes with arbitrary relative amplitudes and radial profiles. In particular, the monopole mode is generically present, so that the slowest decaying contribution of any ``up"--excited $\ell$ mode is generated by the initial monopole. Our proposal degenerates to the Barack--Ori formula when $\ell'=0$. When the monopole is not present in the generic initial data set, our proposal will produce the decay rate of a mixture of modes when $\ell'$ is taken to be the smallest multipole present in the generic data set, and as a particular case will also produce the decay rate of the $\ell$ multipole when the initial data is a pure $\ell' (<\ell)$ mode, the non--generic case, in the language of Barack and Ori. In summary, the Barack--Ori formula Eq.~(\ref{b-o}) is simply the particular case $_0n_{\ell}$ of Eq.~(\ref{prop}). 

Table I summarizes our results for the azimuthal, $m=0$, case. It shows the power law indices for the late time tails for both ``up"- and ``down"-modes. It is in some loose sense a ``partial scattering matrix," that includes (partial) information on the excitation amplitude for each of the excitation channels corresponding with a preparation state of the initial data. The scattering matrix is only partial, because it includes no information on the pre-factor $\psi_0=\psi_0(r_0,\; {\rm profile})$ of the late-time tail 
$\psi (t)\sim \psi_0\,t^{-n}$, where $r_0$ is the evaluation point, and where $\psi_0$ also possibly depends on the radial profile of the initial data. The (partial) scattering matrix includes only the information on $n$. The scattering matrix has a clear shape of the sum of two triangular matrices: letting rows be associated with $\ell'$ and column with $\ell$, the upper triangular matrix elements satisfy $n=\ell'+\ell+3$ and the strictly lower triangular matrix elements satisfy $n=\ell'+\ell+1$.

\subsection*{Non--azimuthal modes}

Our proposal  (\ref{prop}) is not limited to azimuthal modes, and predicts that the late-time decay rate of the $\ell,m$ modes that is excited by the initial $\ell',m$ mode is independent of $m$, and depends only on $\ell,\ell'$. We have tested this hypothesis in numerous cases, that are summarized in Tables I, II, and III. 

\begin{figure}
 \includegraphics[width=3.4in]{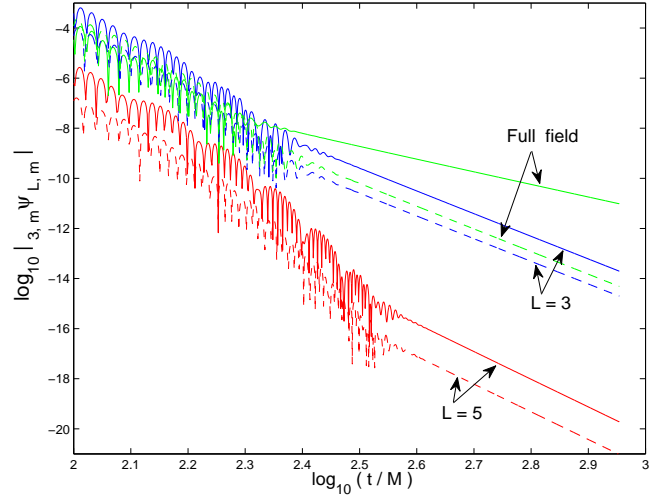} 
\caption{The full fields and their $\ell=3,5$ projections as functions of the time $t$ for $m=1$ (solid) and $m=2$ (dashed). To help identify the modes, we denote the fields with $_{\ell',m}\psi_{\ell,m}$. The initial data are for an initial pure octupole moment. }
\label{tail3m}
\end{figure}

\begin{figure}
 \includegraphics[width=3.4in]{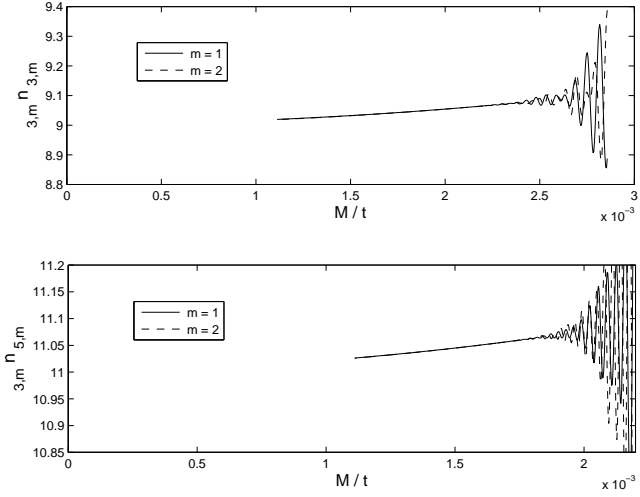} 
\caption{The local power indices $_3n_3$ and $_3n_5$ as functions of the time $t$ for $m=1,2$. To help identify the modes, we denote the local power indices with $_{\ell',m}n_{\ell,m}$. The case $m=1$ is described by a solid curve, and the case $m=2$ is described by the dashed curve.}
\label{lpi3m_bw}
\end{figure}

In Fig.~\ref{tail3m} we show the 
fields for $m=1$ and $m=2$ modes that are created by an initial octupole and excites all odd $\ell$ modes. Specifically, we show the $\ell=3,5$ projections for both $m$ values. We also show the full  fields for both $m$ values. Notably, for the case $m=1$ the dipole mode is also excited, and eventually dominates the full field at late times. In the $m=2$ case the smallest $\ell$-mode that is excited is the octupole, such that the full field in the $m=2$ case decays more rapidly than in the $m=1$ case. We also show the local power indices in Fig.~\ref{lpi3m_bw}. The late time values of the local power indices for each projected $\ell$ mode approach the same values independently of the value of $m$, in accordance with Eq.~(\ref{prop}).

In Table IV we re-organize the data appearing in Tables I,II, and III, so that the $m$-independence of the power-law indices is brought out in a sharper way. Specifically, the late--time decay rates for any allowed excited $\ell'$ mode from an initial $\ell$ mode is independent of the value of $m$ in all the cases we have examined. 

\begin{table}[htdp]
\caption{Scattering matrix for non-azimuthal modes with $m=1$. Late--time power law indices $_{\ell'}n_{\ell}$ for the $\ell$--projections of the fields for pure $m=1$ mode initial data sets, $\ell'=2,3,4,5$. This table includes both ``down" and ``up" excitations. Asterisks relate to kinematically disallowed states. All figures are significant.}
\begin{center}
\begin{tabular}{||c||c|c|c|c|c||}
\hline
Initial  & Projected & Projected & Projected & Projected & Projected \\
$\ell'$ mode & $\ell=1$ & $\ell=2$ & $\ell=3$ & $\ell=4$ & $\ell=5$   \\
\hline
2 & * & 6.997 & * & 8.995 & *  \\
3 &  5.0001 & * & 9.009  &  *  &  11.005   \\
4 & * & 7.0001 & * & 11.006 & *  \\
5& 7.0006 & *  & 9.0001 & * & 13.008  \\
\hline
\end{tabular}
\end{center}
\label{default}
\end{table}%

\begin{table}[htdp]
\caption{Scattering matrix for non-azimuthal modes with $m=2$. Late--time power law indices $_{\ell'}n_{\ell}$ for the $\ell$--projections of the fields for pure $m=2$ mode initial data sets, $\ell'=2,3,4$. This table includes both ``down" and ``up" excitations. Asterisks relate to kinematically disallowed states. All figures are significant.}
\begin{center}
\begin{tabular}{||c||c|c|c|c||}
\hline
Initial  & Projected & Projected & Projected & Projected  \\
$\ell'$ mode & $\ell=2$ & $\ell=3$ & $\ell=4$ & $\ell=5$   \\
\hline
2 & 7.0003 & * & 8.995 & *   \\
3 &  * & 9.0006 & *  &  11.001     \\
4 & 7.0001 & * & 11.003 & *   \\
\hline
\end{tabular}
\end{center}
\label{default}
\end{table}%

\begin{table}[htdp]
\caption{Presentation of the data in Tables I,II, and III in a form that emphasizes the independence of the decay rate on $m$. Late--time power law indices $_{\ell',m}n_{\ell,m}$ for the $\ell$--projections of the fields for pure $m=0,1,2$ mode initial data sets, $\ell=2,3,4$. This table includes both ``down" and ``up" excitations. Asterisks relate to kinematically disallowed states. All figures are significant.}
\begin{center}
\begin{tabular}{||c||c|c|c|c||}
\hline
Initial  & Projected & Projected & Projected & Projected  \\
$\ell',m$ mode & $\ell=2$ & $\ell=3$ & $\ell=4$ & $\ell=5$   \\
\hline
2,0 & 7.009 & * & 9.01 & * \\
2,1 & 6.997 & * & 8.995 & *   \\
2,2 & 7.0003 & * & 8.995 & *   \\
\hline
3,0 & * & 9.009 & * & 11.008 \\
3,1 &  * & 9.009 & *  &  11.005     \\
3,2 &  * & 9.0006 & *  &  11.001     \\
\hline
4,0 & 7.002 & * & 11.008 & * \\
4,1 & 7.0001 & * & 11.006 & *   \\
4,2 & 7.0001 & * & 11.003 & *   \\
\hline
\end{tabular}
\end{center}
\label{default}
\end{table}%

\acknowledgments

We are indebted to Richard Price and Jorge Pullin for discussions. Most of the numerical simulations were executed on Alabama Supercomputer Center (ASC), Louisiana Optical Network Initiative (LONI), and Playstation 3 clusters.

LMB was supported in part by a Theodore Dunham, Jr. Grant of the F.A.R., by a NASA EPSCoR RID grant, and by NSF grants PHY--0757344 and DUE--0941327. GK was supported in part by NSF grants PHY--0831631, PHY--0902026, and PHY--1016906.

\end{document}